\documentclass[twocolumn]{aastex631}

\usepackage{amsmath, amssymb, bm}
\usepackage{graphicx}
\usepackage{multirow}
\usepackage{threeparttable}

\def\aap{Astronomy and Astrophysics}

\def\apjl{ApJL}

\def\apjs{ApJS}

\shorttitle{Constraining GRB Models by GeV Observations}
\shortauthors{Liu et al.}

\begin{document}

\title{Constraints on the Model of Gamma-ray Bursts and Implications from GRB~221009A: \\
GeV gamma rays v.s. High-energy Neutrinos}

\author[0000-0003-1576-0961]{Ruo-Yu Liu}
\affiliation{School of Astronomy and Space Science, Nanjing University, Nanjing 210023, China}
\affiliation{Key laboratory of Modern Astronomy and Astrophysics (Nanjing University), Ministry of Education, Nanjing 210023, China}

\author{Hai-Ming Zhang}
\affiliation{School of Astronomy and Space Science, Nanjing University, Nanjing 210023, China}
\affiliation{Key laboratory of Modern Astronomy and Astrophysics (Nanjing University), Ministry of Education, Nanjing 210023, China}

\author{Xiang-Yu Wang}
\affiliation{School of Astronomy and Space Science, Nanjing University, Nanjing 210023, China}
\affiliation{Key laboratory of Modern Astronomy and Astrophysics (Nanjing University), Ministry of Education, Nanjing 210023, China}

\correspondingauthor{Ruo-Yu Liu}
\email{ryliu@nju.edu.cn}

\begin{abstract}
{Gamma-ray bursts (GRB) are generally believed to be efficient particle accelerators. In the presence of energetic protons in a GRB jet, interactions between these protons and intense radiation field of the GRB are supposed to induce electromagnetic cascade. Electrons/positrons generated in the cascade will produce an additional spectrum of robust feature, which is in the form of a power-law distribution up to GeV regime with an index of $\lesssim 2$. We suggest that measurements of Fermi-LAT at GeV band can provide independent constraints on the key GRB model parameters such as the dissipation radius, the jet's bulk Lorentz factor, and the baryon loading factor. Taking GRB~221009A, the brightest GRB ever detected, as an example, we show that the constraints from GeV gamma-ray emission may be more stringent than that from the neutrino observation, providing us a deep insight into the origin of GRBs.} 
\end{abstract}

\section{Introduction}
Gamma-ray bursts (GRB) are the most violent explosions in the universe. With extreme physical processes and uncommon environments formed during the burst, they are widely considered as accelerators of ultrahigh-energy cosmic rays \citep{Waxman95,Vietri95, Milgrom95}. Accelerated protons (and also heavier nuclei) in a GRB will inevitably interact with its intense radiation field via the photomeson process and Bethe-Heitler process, leading to the production of various high-energy secondary particles such as neutrinos \citep[e.g.,][]{Waxman97, Guetta04, Murase06b} and electromagnetic (EM) particles (i.e., gamma-ray photons and electron/positron pairs). While neutrinos can leave the GRB once they are generated, EM particles would initiate the so-called EM cascade \citep[e.g.,][]{Bottcher98, Dermer06, WangK18}, depositing their energies to lower-energy pairs. The generated pairs will give rise to a broadband radiation, and possibly constitute an extra component of spectrum at GeV band with respect to the main component of the GRB's prompt emission in the keV -- MeV band \citep{Asano09, Asano10, WangK18}.

IceCube neutrino telescope has searched for high-energy neutrinos associated with GRBs, but none of positive detection has been achieved \citep{IC12GRBnu, IC15GRBnu, IC17GRBnu}. The null result poses a strong constraint on the key model parameters of GRBs, such as the energy dissipation radius, the bulk Lorentz factor of the GRB jet and the baryon loading factor of the jet, because the predicted neutrino flux highly depends on these parameters\citep{He12}. On the other hand, the extra spectrum component from the EM cascade is in principle detectable at the GeV band where the main burst component most likely has disappeared. In fact, Fermi-LAT has detected GeV gamma-ray emission from many GRBs during the prompt emission phase \citep{Fermi090902B, Fermi090926A, Tang21}, but the origin of these GeV photons is still unclear yet. The proton-induced EM cascade is one possible interpretation, whereas other processes such as the inverse Compton (IC) radiation of electrons in the prompt emission phase or in the early afterglow phase can also contribute to the observed GeV emission too \citep{Gupta07, Kumar09, Beloborodov14, WangK18, ZhangB22}. Nevertheless, the measured GeV flux (or the upper limits in the case of null detection) in the prompt emission phase can be regarded as an UL for the proton-induced cascade emission, and provide  independent constraints on the GRB models.

GRB~221009A was recently discovered with a record-breaking fluence of roughly $0.052\,\rm erg/cm^2$ in $10-1000\,$keV during its brightest phase between $200\,$s to $600\,$s \citep{Konuswind_GRB221009A} after trigger \citep{GBM_GRB221009A}. As the brightest GRB ever detected, it released an isotropic-equivalent energy of about $3\times 10^{54}\,$erg in the aforementioned energy range given a redshift of $z=0.151$ \citep{redshift_GRB221009A} 
IceCube neutrino telescope found no track-like events from the direction of the GRB in a time range of -1 hour/+2 hours from the trigger time, placing a muon-neutrino upper limit (UL) of $E_\nu^2 dN_\nu/ dE_\nu = 3.9 \times 10^{-2} \rm GeV/cm^2$ at 90\% confidence level (CL), under the assumption of an $E_\nu^{-2}$ power-law spectrum between 0.8\,TeV and 1\,PeV \citep{IC_GRB221009A}. Very recently, some studies \citep{Ai22, Murase22} showed that the neutrino measurement could put useful constraints on the model of this GRB. We will show in this letter that the measurement of Fermi-LAT on this GRB could put even stronger constraints on the model compared to neutrino observations, and has important implications for the origin of GRBs.

\section{Radiation from Electromagentic Cascades Induced by Energetic Protons}
\begin{figure}[htbp]
    \centering
    \includegraphics[width=1\linewidth]{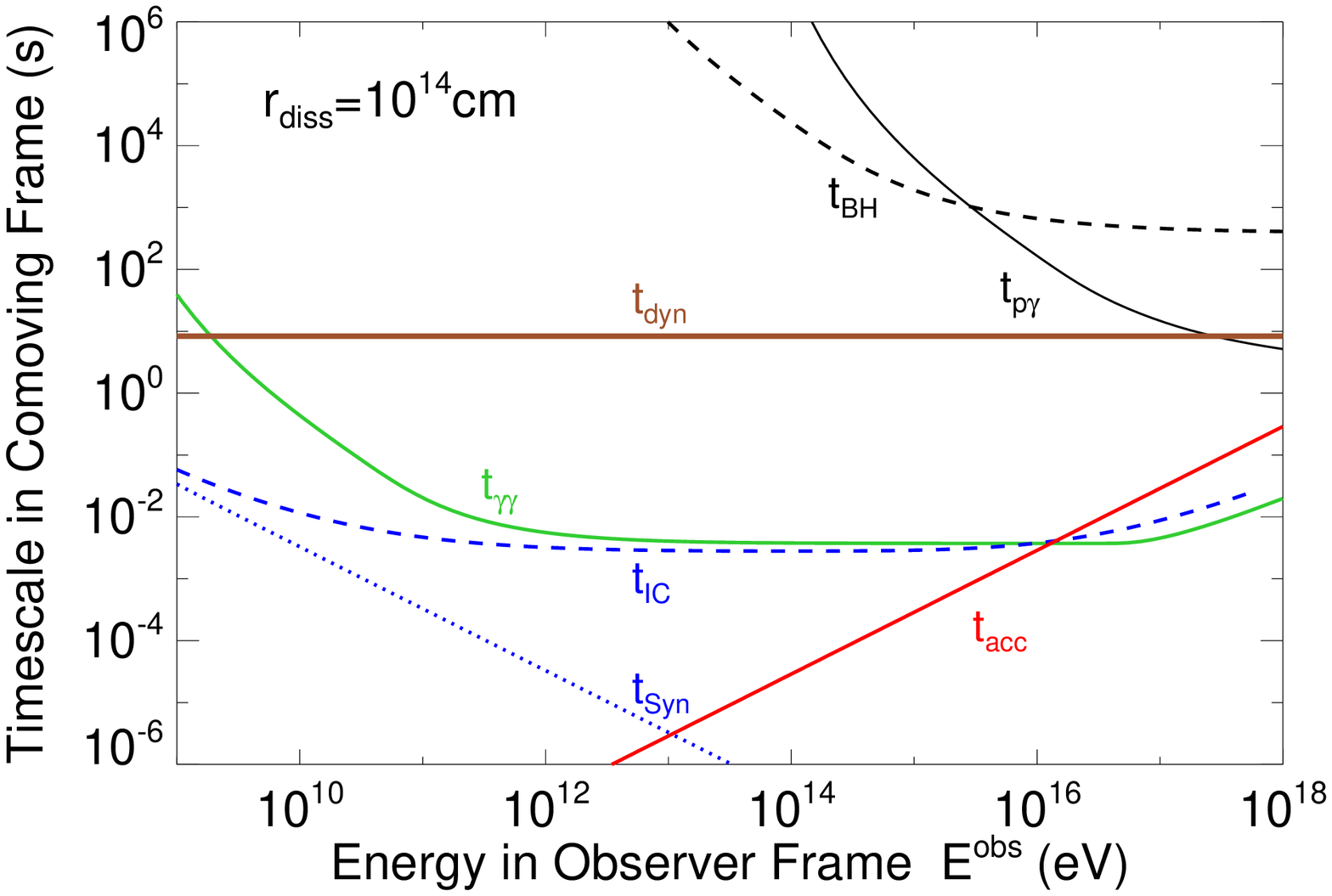}
    \includegraphics[width=1\linewidth]{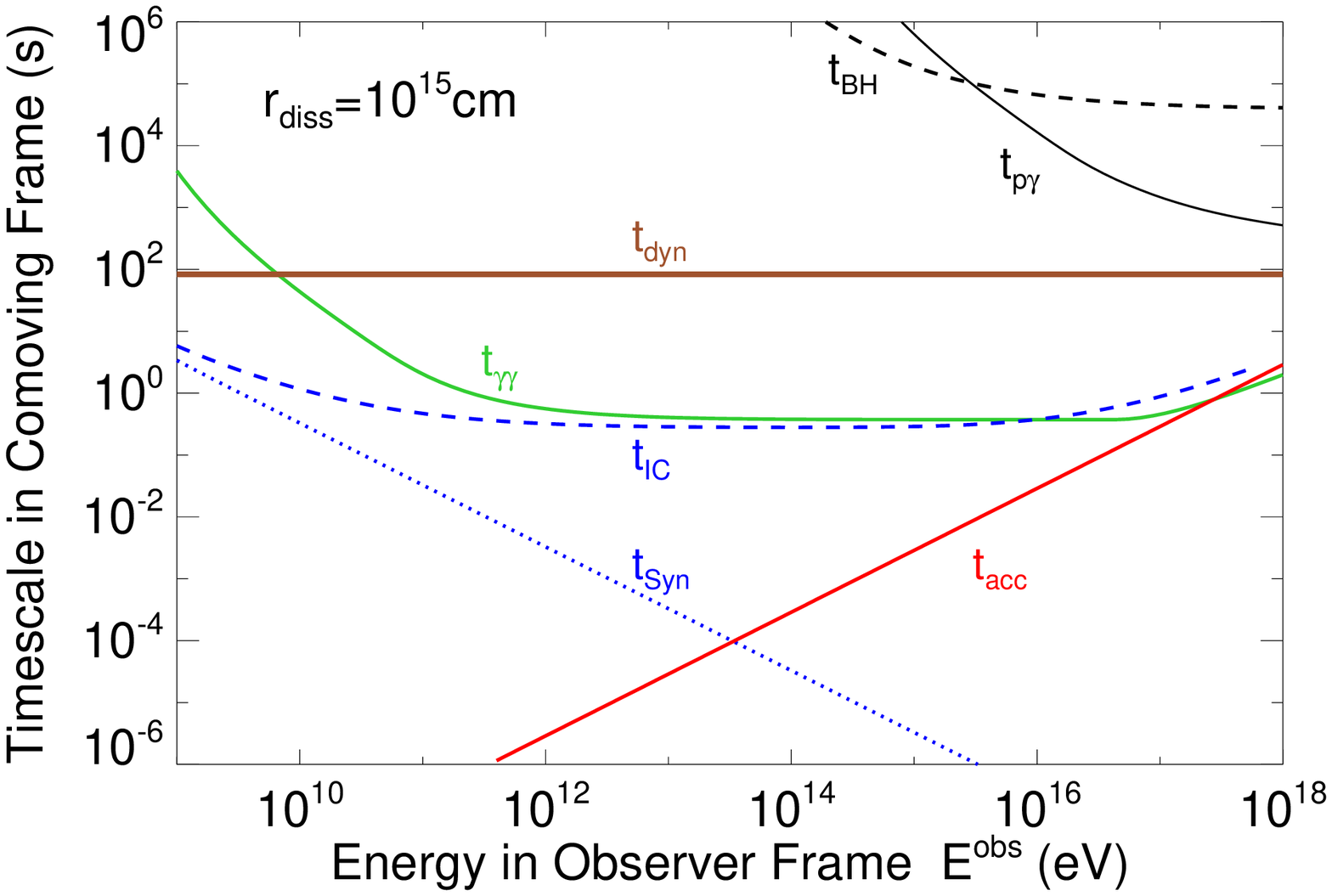}
    \caption{Timescales in the comoving frame of various processes. The solid and dashed black curves represent the energy loss timescales of protons via the photomeson process and BH process respectively. The dashed and dotted blue curves show the cooling timescales of electrons via the IC radiation and synchrotron radiation respectively. The green curve shows the $\gamma\gamma$ absorption timescale of photons. The solid brown curve present the dynamical timescale of the energy dissipation. In both two panels, we assume $L_B=L_{\gamma}=10^{53}\,$erg/s, $\Gamma=400$. The dissipation radius are different in the two panels where $r_{\rm diss}=10^{14}\,$cm for the upper one while $r_{\rm diss}=10^{15}\,$cm for the lower one.}
    \label{fig:timescale}
\end{figure}

Let us start by estimating the proton acceleration in GRB. The most frequently discuss proton acceleration mechanism is the Fermi-type acceleration mechanism (e.g., by shocks) and the acceleration timescale of protons of energy $E_p$ in the comoving frame can be given by $t_{\rm acc}=(20/3)\xi^{-1}E_p/eBc$ \citep{Rieger07}, where $\xi(\leq 1)$ represents the acceleration efficiency, and $B$ is the magnetic field. Assuming energy equipartition between magnetic field and electrons, the magnetic luminosity of the GRB jet is roughly comparable to the bolometric luminosity $L_\gamma$ considering the fast cooling of electrons in the prompt emission phase. We then can estimate the magnetic field in the jet's comoving frame to be $B=\sqrt{2L_\gamma/(r_{\rm diss}^2\Gamma^2c})$ with $r_{\rm diss}$ being the radius of the dissipation radius and $\Gamma$ the bulk Lorentz factor. The maximum proton energy can be estimated by equating the acceleration timescale to the dynamical timescale $t_{\rm dyn}=R/(\Gamma c)$ or the energy loss timescale of protons due to interactions with the radiation field of GRBs via the photomeson process and the Bethe-Heitler (BH) process (see Fig.~\ref{fig:timescale}, where we take $\xi=0.1$). In the former process, a proton can interact with a photon of energy $E_\gamma$ through
\begin{equation}
p+\gamma \to p/n + \left\{
\begin{array}{ll}
\pi^0 \to \gamma+\gamma \\
\pi^+ \to e^+ + \nu_\mu+ \bar{\nu}_\mu + \nu_e
\end{array}
\right.
\end{equation}
if $E_pE_\gamma \gtrsim 0.15\,\rm GeV^2$, while in the latter process we have $p+\gamma \to p+ e^- +e^+$ if $E_pE_\gamma \gtrsim 10^{-3}\,\rm GeV^2$, noting that $E_p$ and $E_\gamma$ here are measured in the comoving frame of the jet. The differential photon density of the GRB's prompt emission is usually described by the Band function \citep{Band93}, which is a smoothly connected broken power-law function and can be characterised with the break energy $E_{\rm b}$, and the low- and high-energy photon indexes $\alpha$ and $\beta$. Denoting the differential photon number density in the comoving frame of the dissipation region by $n_{\rm ph}(E_\gamma)$, the relation between $n_{\rm ph}$ and $L_\gamma$ can be given by $\int n_{\rm ph}E_\gamma dE_\gamma = L_\gamma/4\pi \Gamma^2 r_{\rm diss}^2 c$. 

Neutrinos generated in the photomeson process will leave the dissipation region directly. However, EM particles will interact with the radiation field and the magnetic field of the GRB: high-energy photons will be absorbed or annihilated with low-energy photons into pairs, and pairs, which are generated either in the annihilation, the photomeson process or the BH process, will radiate via the synchrotron radiation and the inverse-Compton (IC) scattering off the GRB's radiation field. High-energy photons radiated by pairs will be annihilated again as long as their energies is sufficiently high. Such a cycle of $e-\gamma$ conversion will be repeated many times, generating more and more low-energy secondary EM particles until the energies of newly generated photons drop below a critical energy at which the opacity of $\gamma\gamma$ annihilation is equal to unity. In Fig.~\ref{fig:timescale}, we show the gamma-ray annihilation timescale in the jet's comoving frame compared with the dynamical timescale. The cooling timescales of electrons/positrons are also shown, where we see that the synchrotron radiation dominates over the IC radiation because the latter suffers the Klein-Nishina effect. It is important to note that the radiative cooling timescales are shorter than the dynamical timescale at high energies. We therefore may regard the EM cascade process has reached the quasi-steady state, and calculate its radiation following the method detailed in \citet{Liu20_tde}.

\begin{figure}[htbp]
    \centering
    \includegraphics[width=1\linewidth]{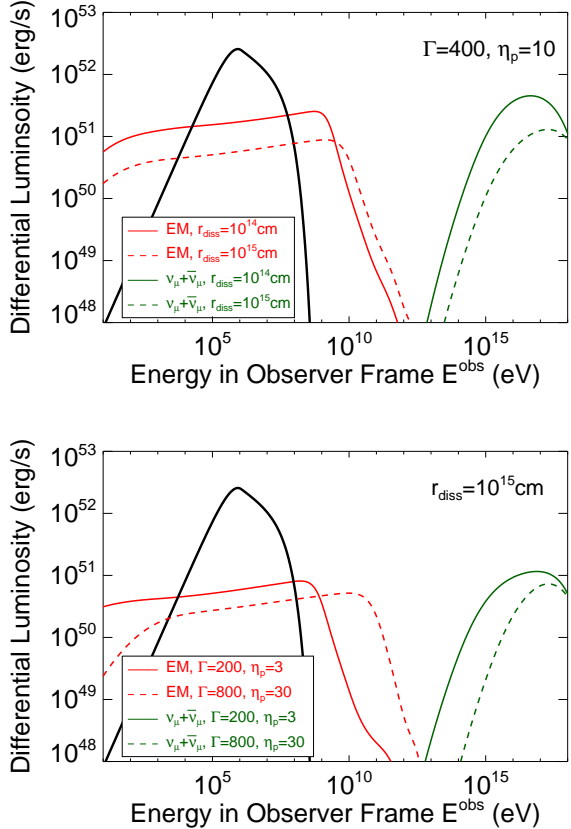}
    \caption{Predicted SEDs of the EM cascade emission (red curves) and the high-energy neutrino emission (green curves). In the upper panel, we set $\Gamma=400$ and $\eta_p=10$. Solid curve and dashed curve are for $r_{\rm diss}=10^{14}\,$cm and $10^{15}$\,cm respectively. In the lower panel, we set $r_{\rm diss}=10^{15}\,$cm. Solid curves show the result for $\Gamma=200$, $\eta_p=3$ while dashed curves for $\Gamma=800$, $\eta_p=30$. In both  panels, the black curves represent the SED of the Band component, with $\alpha=-1$, $\beta=-2.3$ and $E_{\rm peak}^{\rm obs}=1\,$MeV. $L_B=L_\gamma=10^{53}\rm erg/s$ is fixed.}
    \label{fig:spectrum}
\end{figure}

In Fig.~\ref{fig:spectrum}, we show the spectra of the radiation generated in the proton-induced EM cascade and the accompanying (anti-)muon neutrino for different model parameters. In both  panels of the figure, we assume the luminosity of the prompt emission (i.e., the Band component) in 10\,keV -- 10\,MeV to be $L_\gamma=10^{53}\, $erg/s, with a Band function of $\alpha=-1$, $\beta=-2.3$ and $E_{\rm b}^{\rm obs}=1\,$MeV followed by an exponential cutoff at 50\,MeV. The spectral shapes of the cascade radiation in general appear similar among different cases. They all approximately follow a power-law function with index $\lesssim 2$ up to the GeV regime. Such a universal spectrum can be formed as long as the cascade has been developed sufficiently, i.e., $t_{\rm syn}$ or $t_{\rm IC} \ll t_{\rm dyn}$ (in other words, the quasi-steady state is established). Under this condition, the electrons/positrons in the cascade are mainly created from the annihilation of gamma rays, so their spectrum at generation basically follows that of the gamma rays spectrum. However, due to the rapid radiative cooling dominated by the synchrotron radiation, the electron/positron spectrum will be steepened in the quasi-steady state with the spectral index being increased by 1. We denote the photon index of the gamma-ray spectrum by $s_\gamma$ and the spectral index of pairs in the quasi-steady state by $s_e$, their relation is given by $s_e= s_\gamma + 1$. On the other hand, the gamma-ray photons are mainly emitted via the synchrotron radiation in the cascade, so we have $s_\gamma=(s_e+1)/2$. These two relations between $s_\gamma$ and $s_e$ would give us $s_e=3$ and $s_\gamma=2$, which is basically consistent with what we show in Fig.~\ref{fig:spectrum}. The EM cascade spectrum cannot maintain a power-law behaviour to very high energies but will be truncated by the $\gamma\gamma$ annihilation at certain energy around $0.1-100\,$GeV in the four cases shown, depending on where the opacity of the annihilation, which can be estimated by $t_{\rm dyn}/t_{\gamma\gamma}$, reaches unity. A larger dissipation radius and/or a higher bulk Lorentz factor lead to a weaker gamma-ray absorption and hence results in a higher truncation energy.

We also see that the flux level of (anti-)muon neutrino and the EM cascade is always comparable. This can be roughly understood as follows. Half of the proton energy lost in the photomeson process goes into $\pi^+$, from which three neutrinos and one positron are created. The other half of the lost proton energy goes into $\pi^0$ and further decays into photons. As a result, (anti-)muon neutrinos take away $1/8$ of the proton energies lost in photomeson process after considering the neutrino flavor oscillation, while EM particles carry $5/8$ of the lost energies. On the other hand, the cascade process re-distributes the energy of the first-generation EM products over a broad range of about $6-7$ orders of magnitude with a roughly flat spectrum, so that the EM flux at each logarithmic interval need be divided by $\ln 10^6\sim 10$ and finally becomes a few times lower than that of the neutrino flux. 


\section{Application to GRB~221009A}

\subsection{Observations of GRB~221009A by Fermi}

GRB 221009A triggered Fermi-GBM at 
13:16:59.99 UT ($\rm T_0$) on 2022 October 9 \citep{GBM_GRB221009A}.
The GRB was also detected by Fermi-LAT after 200\,s of the Fermi-GBM trigger, despite a large angle ($>70^\circ$) from the LAT boresight at this time \citep{LAT_GRB221009A}.

\begin{figure*}[htbp]
    \centering
    \includegraphics[width=0.8\textwidth]{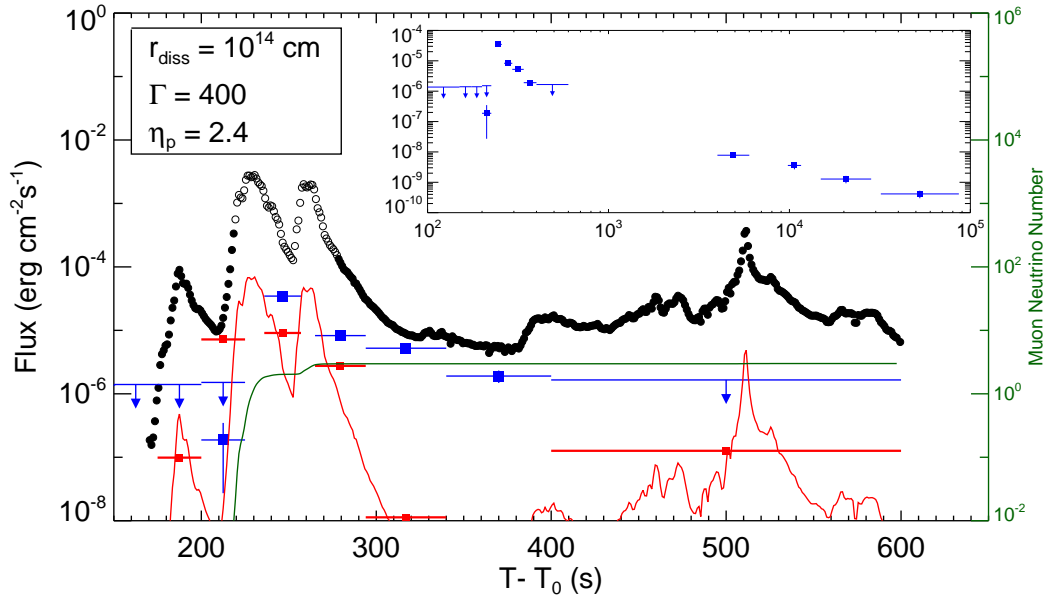}
    \caption{$0.1-10$\,GeV lightcurve observed by Fermi-LAT (blue squares or ULs) in comparison with the prediction from the proton-induced cascade emission in the same energy range (the red curve) with $r_{\rm diss}=10^{14}\,$cm and $\Gamma=400$. Red horizontal bars shows the predicted average flux in each time bin of Fermi-LAT data. The gaps in $225-236\,$s and $257-266\,$s are due to the instrumental pile-up effects and hence the data is dropped. The green curve represent the expected cumulative $\nu_\mu+\bar{\nu}_\mu$ event number in IceCube, which is normalized to 3 by adjusting the baryon loading factor to be $\eta_p=2.4$. Black circles show the 10\,keV -- 10\,MeV lightcurve measured by Fermi-GBM, where the open circles represent the data that may be under the influence of the pile-up effects. For reference, the inset presents the complete $0.1-10\,$GeV lightcurve observed by Fermi-LAT up to one day after the trigger, covering the afterglow phase.}
    \label{fig:lc}
\end{figure*}

GBM has 12 sodium iodide (NaI) and two bismuth germanate (BGO) scintillation detectors, covering the energy range 8 keV$-$40 MeV \citep{Meegan09}. 
We downloaded GBM data of this GRB from the Fermi-GBM public data archive.\footnote{\url{https://heasarc.gsfc.nasa.gov/FTP/fermi/data/gbm/daily/}}
For this burst, the detectors selected for our analysis are two NaI detectors (namely $n7$ and $n8$) and one BGO detector (namely b1), which have the smallest viewing angles with respect to the GRB source direction. The light curve of GRB 221009A in $\rm 10~keV-10~MeV$, as observed with GBM is shown in Fig. \ref{fig:lc}. The light curve is obtained by assuming the Band function for the spectrum in a 1.024~s time-bin. Due to the extraordinary brightness of this GRB, the pile-up instrumental effect makes the spectrum and flux between $T_0+219$ and $T_0+278$ uncertain, but this would not change our result and conclusion dramatically. Details of the analysis of Fermi-GBM data is shown in Appendix.~\ref{app:A}. 

The Fermi-LAT extended type data for the GRB 221009A were taken from the Fermi Science Support Center\footnote{\url{https://fermi.gsfc.nasa.gov}}. 
Only the data within a $14\degr \times14\degr$ region of interest (ROI) centered on the position of GRB 221009A are considered for the analysis. The publicly available Pass 8 (P8R3) LAT data for GRB 221009A was processed using the Conda fermitools v2.2.0 package, distributed by the Fermi Collaboration. Events of the ``Transient'' class ($P8R3\_TRANSIENT020\_V3$; using for the time before $\rm T_0+400$\,s) and "Source" class ($P8R3\_SOURCE\_V3$; using for the time after $\rm T_0+400$\,s) were selected. We assumed a power-law spectrum in the $0.1-500$ GeV energy range, with accounting for the diffuse Galactic and extragalactic backgrounds. We did not use the data in the time intervals $\rm T_0+225~s$ to $\rm T_0+236~s$ and $\rm T_0+257~s$ to $\rm T_0+265~s$ considering the pile-up effect due to the extremely high flux at the time \citep{LAT_221009A_pileup}. 

The Fermi-LAT light curve of GRB 221009A in $\rm 0.1~GeV-10~GeV$ is shown in Fig.~\ref{fig:lc}. Significant detection of Fermi-LAT on this GRB starts around $T_0+200\,$s, with 3 photons recorded in the time interval of $T_0+200\,$s and $T_0+225\,$s with $\rm TS=10$, leading to an average energy flux of $1.9\times 10^{-7}\rm erg/cm^2s$ in this interval. \citet{LATBTI_GRB221009A} further extended the Bad Time Interval (BTI) to $\rm T_0+203~s$ and $\rm T_0+294~s$ because of the X-ray and soft gamma-ray pile-up in the anticoincidence detector (ACD). The additional activity of the ACD can lead to misclassification of gamma rays as background. As a result, the flux during the BTI may be underestimated. For the sake of conservation, we choose a strict criteria for detection (i.e., $\rm TS\geq25$) in this time interval, resulting in a 95\% C.L. UL of $1.5\times10^{-6}\rm erg/cm^2/s$ in $0.1-10$\,GeV. We will use this UL instead of the aforementioned flux in our calculation later. See Appendix.~\ref{app:B} for more details about Fermi-LAT data analysis.


\subsection{Result}
Although the measured spectrum of the main component varies with time, we assume a time-independent spectral shape of Band function for the main component same as the one shown in Fig~\ref{fig:spectrum} for simplicity. By normalizing the 10\,keV -- 10\,MeV flux to the GBM observation at each time, we can calculate the lightcurve of the EM cascade emission in 0.1 -- 10\,GeV as well as that of the high-energy neutrino emission. We require the predicted 0.1 -- 10\,GeV flux at any time bin not to overshoot the flux or the 95\% C.L. UL of Fermi-LAT, so that upper limits of baryon loading factor can be constrained for any given dissipation radius $r_{\rm diss}$ and the bulk Lorentz factor $\Gamma$. On the other hand, IceCube's non-detection of (anti-)muon neutrino events from this GRB provides an independent constraint. However, we note that the UL of the time-integrated neutrino fluence given by IceCube, i.e., $3.9\times 10^{-2}\rm GeV/cm^2$ between 1\,TeV -- 1\,PeV, is obtained assuming an $E_\nu^{-2}$ spectrum. The assumed neutrino spectral shape is inconsistent with the theoretical one predicted for this GRB as shown in Fig.~\ref{fig:spectrum}, which is also pointed by \citet{Murase22}. Therefore, we instead use the fact that no track-like event is detected during the GRB as a constraint for the neutrino emission. This condition can be translated to the maximum expected event number of $\nu_\mu+\bar{\nu}_\mu$ in the range of 1\,TeV -- 1\,EeV, denoted by $N_{\nu_\mu}$, based on the predicted neutrino flux and IceCube's effective area for a point-lie source at the declination of this GRB (i.e., the one for $\delta=-5^\circ -30^\circ$ \citealt{IC20_ps}). The 95\% C.L. UL for the neutrino flux can be obtained by setting $N_{\nu_\mu}=3$, since the probability of non-detection will be less than 5\% for $N_{\nu_\mu}>3$ given that the detection probability follows the Poisson distribution. This can also give us an UL for $\eta_p$.

In Fig.~\ref{fig:lc}, we compare the 0.1 -- 10\,GeV flux of GRB~221009A measured by Fermi-LAT (blue symbols) and the predicted lightcurve (the red curve), assuming $r_{\rm diss}=10^{14}\,$cm and $\Gamma=400$. The predicted lightcurve for the GeV emission basically follows the trend of the  keV -- MeV emission, but the amplitude is not linearly scaled. This is because both the proton injection luminosity and the interaction efficiency are proportional to the keV -- MeV luminosity. The cumulative $\nu_\mu+\bar{\nu}_\mu$ event number expected in IceCube is also shown (the dark green curve, corresponding to the vertical axis on the right) whereas the event number at the end of the prompt emission is normalized to 3 by adjusting the value of the baryon loading factor to $\eta_p=2.4$. In this case, we observe that the expected GeV flux exceeds the UL of Fermi-LAT in 200 -- 225\,s by almost one order of magnitude. It implies that the GeV gamma-ray observation in this time bin gives a stronger constraint on the baryon loading factor $\eta_p$ than that given by the neutrino observation. 

\begin{figure*}[htbp]
    \centering
    \includegraphics[width=0.8\textwidth]{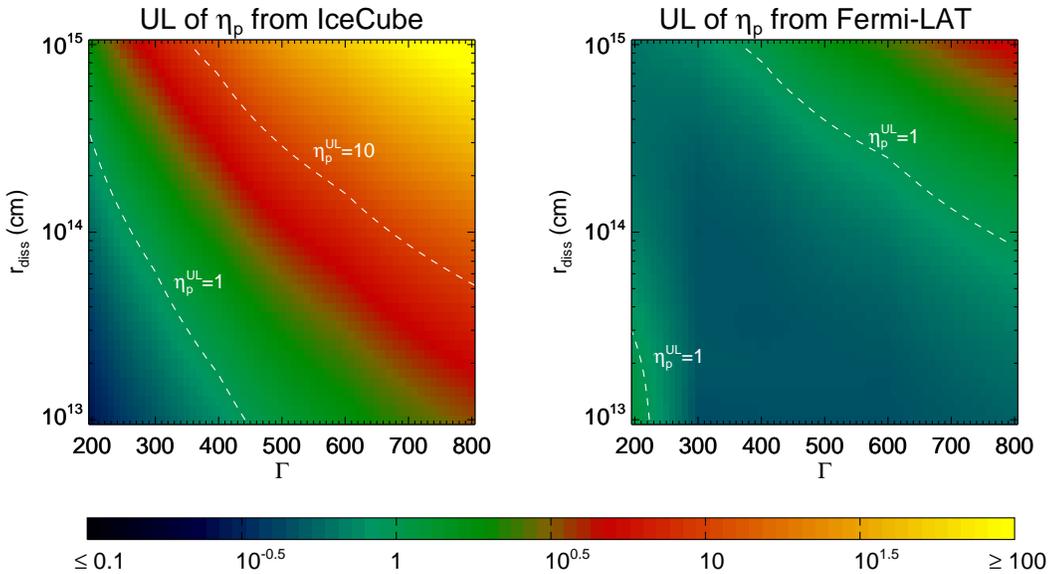}
    \caption{Upper limits of the baryon loading factor $\eta_p$ at different $r_{\rm diss}$ and $\Gamma$. The left panel shows the constraint from the IceCube's observation while the right panel shows the constraint from the Fermi-LAT's measurement.}
    \label{fig:constraint}
\end{figure*}

We repeat the aforementioned calculation for the lightcurves of the neutrino flux and the GeV gamma-ray flux for different combinations of $\Gamma$ and $r_{\rm diss}$, and put the constraint on the baryon loading factor $\eta_p$ according to the IceCube observation and the Fermi-LAT observation respectively. The results are shown in Fig.~\ref{fig:constraint}. Apparently, the constraint from Fermi-LAT (the right panel) is generally much more restrictive than that from IceCube (the left panel) except for very small dissipation radii $r_{\rm diss}$ with a relatively low bulk Lorentz factor $\Gamma$ of the GRB jet. The constraint from IceCube becomes stronger for smaller $r_{\rm diss}$ and $\Gamma$, where the pion production efficiency is high. On the other hand, the constraint from Fermi-LAT does not show a clear dependence on these two parameters. This is due to the effect of the $\gamma\gamma$ annihilation. Although the production efficiency of EM particles would be higher for smaller $r_{\rm diss}$ and $\Gamma$, the $\gamma\gamma$ absorption become stronger in the mean time. These two effects cancel the influences of each other on the GeV gamma-ray flux to certain extent, and lead to a roughly model-independent cascade flux in $0.1-10\,$GeV. Such a phenomena can be also seen straightforwardly from Fig.~\ref{fig:spectrum}: the cascade flux in the case of $\Gamma=200$ and $\eta_p=3$ is generally higher than the case of $\Gamma=800$ and $\eta_p=30$, but the integrated flux in $0.1-10\,$GeV of the latter case is comparable to that of the former case, because the spectrum is truncated before 1\,GeV in the former case. As a consequence, the UL of $\eta_p$ obtained from the Fermi-LAT observation is concentrated in a narrow range $0.5\lesssim \eta_p^{\rm UL} \lesssim 1$ for a large area in the parameter space. The $\gamma\gamma$ absorption on the gamma-ray spectrum also explains the reason why the constraint from the neutrino observation is stronger than that from the GeV gamma-ray observation for $r_{\rm diss}\lesssim 10^{14}\,$cm and $\Gamma\lesssim 300$, where GeV gamma rays are severely absorbed in the dissipation region but neutrinos can escape.

The small value for the UL of $\eta_p$ obtained from the Fermi-LAT observation implies a few interesting possibilities. First, the GRB jet may be probably dominated by leptons and magnetic field. Otherwise, if an efficient baryon loading procedure presents in the GRB jet with $\eta_p>10$, for example, it would then require a large dissipation radius and a high bulk Lorentz factor, i.e., $r_{\rm diss}>10^{15}\,$cm and $\Gamma >800$. Both these two possibilities would favor the Internal Collision-induced Magnetic Reconnection and Turbulence (ICMART) Model \citep{Zhang11}, in which the GRB jet is dominated by the Poynting flux and the energy dissipation occurs at a large radius of $10^{15}-10^{16}\,$cm. Alternatively, the strong constraint on $\eta_p$ may be circumvented by arguing that GRB is not an efficient proton accelerator, with either a low acceleration efficiency (i.e., a small maximum proton energy) or producing a quite soft proton spectrum. In this case, we may rule out GRBs as the sources of ultrahigh-energy cosmic rays.

\section{Conclusion}
In summary, if accelerations of energetic protons (or nuclei) take place in GRBs, these energetic particles will inevitably interact with the intense radiation of GRBs in keV -- MeV band via the photomeson process and the BH process. The interactions will yield high-energy neutrinos and induce electromagnetic cascades which would finally form an additional radiation component with spectrum continuing to GeV gamma-ray band. Observations by IceCube and Fermi-LAT on GRBs can thus set an upper limit, respectively, for the theoretical neutrino flux and GeV gamma-ray flux. The upper limits can be translated to the constraints on the key model parameters of GRBs such as the energy dissipation radius $r_{\rm diss}$, the bulk Lorentz factor of the GRB jet $\Gamma$, and the baryon loading factor $\eta_p$. We analyzed the Fermi-LAT observations on the brightest-ever GRB~221009A and calculated the high-energy neutrino flux and the radiation of the electromagnetic cascade under different combinations of those key model parameters. By comparing the predicted neutrino flux and GeV gamma-ray flux with the measurement of IceCube and Fermi-LAT on GRB~221009A respectively, we found that the constraint from the GeV observation are stronger than that from neutrino observations. More specifically, for a large area in the $\Gamma-r_{\rm diss}$ space, we obtained a quite stringent upper limit of the baryon loading factor to be $0.5\lesssim \eta_p^{\rm UL}\lesssim 1$. This result may imply that the GRB jet is dominated by leptons and magnetic field or otherwise requires a large dissipation radius and jet's bulk Lorentz factor. Alternatively, considering a soft proton spectrum or a small maximum energy in the proton spectrum would relax the constraint, but this would disfavor GRBs as the main accelerators of ultrahigh-energy cosmic rays.

Finally, we caveat again about the pile-up effect on the Fermi data. We found that the most constraining Fermi-LAT data point is in the time interval between $T_0+200\,$s and $T_0+225\,$s. The LAT flux at this time bin may be underestimated due to the potential pile-up in ACD. Although we used a conservative upper limit, which is about 8 times that of the measured flux, in constraining the baryon loading factor, the true GeV flux at this time interval might be even higher than this conservative upper limit. If so, the obtained constraint would be relaxed to certain extent, depending the true GeV flux level. In the worst case, we could use the Fermi-LAT upper limit at the time interval in ($\rm T_0+175\,$s) -- ($\rm T_0+200\,$s) or in ($\rm T_0+400\,$s) -- ($\rm T_0+600\,$s) to constrain the GRB model although the obtained constraints would be weaker than that from the neutrino observation. Nevertheless, regardless of the uncertainty caused by the instrumental effect, we propose that the GeV observation provides an independent way of constraining GRB models, and can be applied to other GRBs in particular those with high keV -- MeV flux.

\appendix
\section{Details of Fermi-GBM Data Analysis}\label{app:A}

The GBM detectors collected data in two different types: temporally pre-binned (CTIME and CSPEC) or temporally unbinned (TTE) data. 
 We downloaded GBM data of this GRB from the Ferm-GBM public data archive.\footnote{\url{https://heasarc.gsfc.nasa.gov/FTP/fermi/data/gbm/daily/}}
 Firstly, we estimated the time interval of the pile-up effect using the TTE data, and we found that the count rate become pile-up start at $\sim \rm T_0+219~s$ and end at $\sim \rm T_0+278~s$.
Then, we extracted the light curves and performed spectral analysis based on the GBM Data Tools (\textit{gbm} package) with the CSPEC data. The light curve in $\rm 10~keV-10~MeV$ was shown in Fig. \ref{fig:lc}, and it was obtained by assuming the Band function for the spectrum in a 1.024~s time-bin. For the pile-up interval $\rm T_0+219~s$ to $\rm T_0+278~s$, the light curve was obtained by assuming the Band function with the typical value $\alpha=-1.0$ and $\beta=-2.3$.

\begin{figure*}[htbp]
    \centering
    \includegraphics[angle=0,scale=0.45]{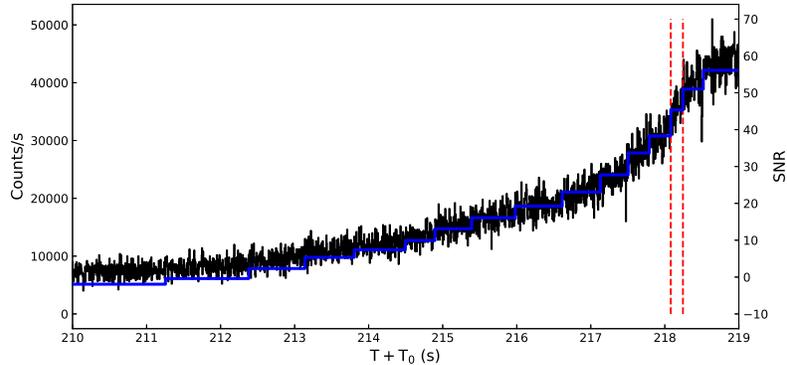}
    \caption{The background subtracted light curves in $8.0-900.0$\,keV extracted from the TTE data, which use a 5 ms time bin. The blue solid line represents the Bayesian block light curve and the red dashed lines represent the minimum bin size of the obtained block.}
    \label{fig:lc_gbm}
\end{figure*}

To determine the variability time scale of this GRB, we employ the Bayesian block method \citep{2013ApJ...764..167S} on the TTE data in 5~ms time-bin (but not including the pile-up interval). The minimum bin size of the obtained blocks is 165 ms and take half of the minimum bin size as the variability time scale. As shown in Fig. \ref{fig:lc_gbm}, our result yields $\tau_{\rm var}\sim82 \rm ms$ for this GRB, and implying a maximum dissipation radius of $r_{\rm diff}=2\Gamma^2c\tau_{\rm var}=8\times 10^{14}(\Gamma/400)^2\,$cm, which is generally consistent with the range of $r_{\rm diff}$ we explored in this study.

\section{Details of Fermi-LAT Data Analysis}\label{app:B}

We perform an unbinned maximum likelihood analysis, using LAT $TRANSIENT$ and $SOURCE$ events between 100 MeV and 500 GeV, and with a maximum zenith angle of 100$\degr$ to reduce the contamination from the $\gamma$-ray Earth limb.
We select a time interval of 0--600 s after the GBM trigger time $\rm T_0$, which contains all the gamma-rays detected by LAT before the GRB exited its field of view (FOV). 
The instrument response function (IRF) (\textit{$P8R3\_TRANSIENT020\_V3$})\footnote{\url{https://fermi.gsfc.nasa.gov/ssc/data/analysis/documentation/Cicerone/Cicerone_Data/LAT_DP.html}} is used.
The main background component consists of charged particles that are misclassified as gamma-rays. It is included in the analysis using the isotropic emission template (``$iso\_P8R3\_TRANSIENT020\_V3\_v1.txt$'').

Due to the GRB second time turn into its FOV is $\sim \rm T_0+4000~s$, the IRF (\textit{$P8R3\_SOURCE\_V3$}) and the corresponding isotropic emission template (``$iso\_P8R3\_SOURCE\_V3\_v1.txt$'') are used. Also, we consider all the Fourth Fermi-LAT source catalog sources \citep{2020ApJS..247...33A} within $10\degr$ centre on the GRB.
The contribution from the Galactic diffuse emissions is accounted for by using the diffuse Galactic interstellar emission template (IEM; $gll\_iem\_v07.fits$). The parameter of isotropic emission and IEM are left free.

\begin{figure*}[htbp]
    \centering
    \includegraphics[angle=0,scale=0.7]{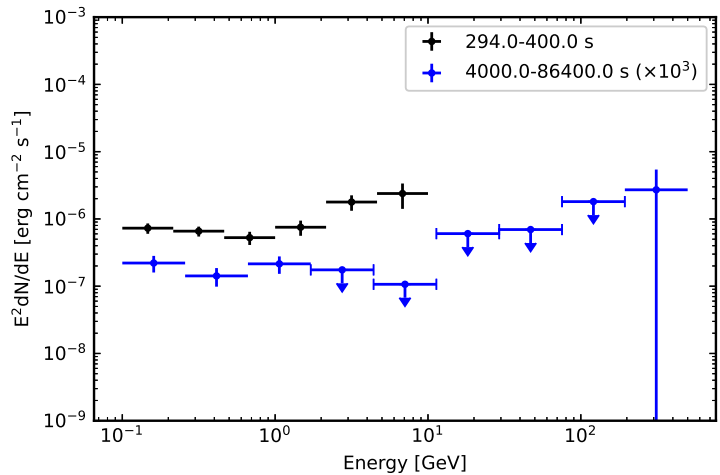}
    \caption{The measured spectrum of LAT for GRB 221009A in 294--400 s (black points) and 4000.0--86400.0 s (blue points). The latter is multiplied by a factor of 1000 for visibility.}
    \label{sed:lat}
\end{figure*}

The maximum likelihood test statistic (TS) is used to estimate the significance of the GRB, which is defined by TS$= 2 (\ln\mathcal{L}_{1}-\ln\mathcal{L}_{0})$, where $\mathcal{L}_{1}$ is maximum likelihood value for the template including the GRB
and $\mathcal{L}_{0}$ is the maximum likelihood value without the GRB (null hypothesis).
The TS value, spectral index and the corresponding flux of different intervals for GRB 221009A are shown in Table \ref{tab:lat}.
The measured spectrum of the GeV emission for GRB 221009A is shown in Fig.~\ref{sed:lat} for reference.

\begin{table}[ht!]
\centering
\caption{Results of the LAT analysis for the GRB 221009A.}
\begin{threeparttable}
    \begin{tabular}{lcccc}
        \hline\hline 
Time intervals & TS\tnote{a} & $\rm Flux_{0.1-10~GeV}$ & Index\tnote{b} & \\
 &  & $\rm erg~cm^{-2}~s^{-1}$&  & \\
\hline
0.0--50.0 &  3.52 & $1.35\times10^{-6}$ & $2.0^*$ & \\
50.0--100.0 & $\sim0.0$ & $1.36\times10^{-6}$ & $2.0^*$ & \\
100.0--150.0 & $\sim0.0$ & $1.36\times10^{-6}$ & $2.0^*$ & \\
150.0--175.0 & $\sim0.0$ & $1.40\times10^{-6}$ & $2.0^*$ & \\
175.0--200.0 & $\sim0.0$ & $1.40\times10^{-6}$ & $2.0^*$ & \\
200.0--225.0 & 10.69 & $(1.87\pm1.60)\times10^{-7}$ & $2.76\pm1.37$ & \\
236.0--257.0 & 3777.46 & $(3.47\pm0.32)\times10^{-5}$ & $1.84\pm0.07$ & \\
265.0--294.0 & 899.09 &  $(8.27\pm1.00)\times10^{-6}$ & $2.03\pm0.10$ & \\
294.0--340.0 & 700.13 & $(5.23\pm0.72)\times10^{-6}$ & $1.71\pm0.09$ & \\
340.0--400.0 & 172.59 & $(1.91\pm0.42)\times10^{-6}$ & $2.09\pm0.18$ & \\
400.0--600.0 & $\sim0.0$ & $1.65\times10^{-6}$ & $2.0^*$ & \\
4000.0--6000.0 & 182.87 & $(7.78\pm1.36)\times10^{-9}$ & $2.07\pm0.13$ & \\
9800.0--11600.0 &  77.15 & $(3.59\pm0.92)\times10^{-9}$ & $2.16\pm0.20$ & \\
14900.0--28300.0 & 49.05 &  $(1.28\pm0.35)\times10^{-9}$ & $2.23\pm0.24$ & \\
32000.0--86400.0 & 34.78 & $(4.14\pm1.15)\times10^{-10}$ & $2.25\pm0.24$ & \\
\hline\hline
    \end{tabular}
 \begin{tablenotes}
        \footnotesize
        \item[a] TS value of each interval, the significance of the GRB is approximate to $\rm \sqrt{TS} \ \sigma$, the TS value of interval which  less than 9 will be estimate as a 95\% C.L. UL. 
        \item[b] The photon index. ULs are calculated with a photon index with $\alpha=2.0$ (labeled with $^*$).
 \end{tablenotes}

\label{tab:lat}
\end{threeparttable}
\end{table}{}

\bibliography{ms}

\end{document}